\documentstyle[epsf]{article}
\huge
 
\def\setzero{\setcounter{equation}{0}}

\newcounter{eqalph}

\begin{document}

\baselineskip 18pt

\def \sech{{\rm sech}}
\def \tanh{{\rm tanh}}
\def \cn{{\rm cn}}
\def \sn{{\rm sn}}
\def\bm#1{\mbox{\boldmath $#1$}}
\newfont{\bg}{cmr10 scaled\magstep4}
\newcommand{\bzr}{\smash{\hbox{\bg 0}}}
\newcommand{\bzl}{%
   \smash{\lower1.7ex\hbox{\bg 0}}}
\title{Unitary Matrix Models  with  a topological term
   \\    and  \\discrete time Toda equation} 
\date{\today}
\author{ Masato {\sc Hisakado}  
\\
\bigskip
\\
{\small\it Graduate School of Mathematical Sciences,}\\
{\small\it University of Tokyo,}\\
{\small\it 3-8-1 Komaba, Megro-ku, Tokyo, 113, Japan}}
\maketitle

\vspace{20 mm}

Abstract

We study  the  full unitary matrix models.
Introducing  a new term $l\log U$, 
 $l$  plays  the role of  the discrete time.
On the other hand,
 the full unitary matrix model 
 contains a topological term.
In the continuous limit
 it gives rise to a phase transition 
at $\theta=\pi$.
The ground state  is characterize 
 by the discrete time $l$.
The discrete time $l$ plays like 
the instanton number.

\newpage

\section{Introduction}

Models  of the symmetric unitary matrix model are 
solved exactly in the double scaling limit,
 using orthogonal polynomials 
on a circle.\cite{p}
The partition function is the form
$\int dU\exp\{-\frac{N}{\lambda}{\rm tr} V(U)\}$, 
where $U$ is an $N\times N$ unitary matrix 
and tr$V(U)$ is some well defined function of $U$.
When $V(U)$ is the self adjoint  we call the model symmetric.\cite{b}
The simplest case is given by $V(U)=U+U^{\dag}$.
This unitary models has been studied in 
connection with the large-$N$ approximation 
to QCD in two dimensions.({\it one-plaquette model})\cite{gw}
The full non-reduced unitary model was first discussed 
in \cite{m2}.
The simplest  potential  is divided into the symmetric and 
anti-symmetric part.\cite{h2}
The symmetric part is the usual Wilson action.
The anti-symmetric part becomes the topological term.\cite{kts}
In the continuous limit we  can reconstruct
the topological information
encoded in the theta term.
We introduce a new term $l\log U$.
The full unitary matrix model can be embedded 
 in the two-dimensional Toda lattice hierarchy.\cite{m2}
Adding the new term to the full unitary matrix model,
 $l$ becomes the discrete time  and 
the partition function  satisfies 
the discrete time Toda equation.
This equation is sometimes called as ``Hirota equation''.\cite{w}
Discrete time integrable system is currently under 
extensive study.
The partition function for the Hermite matrix model with 
the new term  also satisfy the discrete time Toda equation. 
\cite{h}
In the full unitary matrix model 
 the discrete time $l$ couples to the theta  term and characterizes
the ground state.
The partition function for  the $SU(N)$ case  is the 
sum of  the partition function for  the $U(N)$ case 
about all the discrete time $l$.
Thus the theta term would have been zero for $SU(N)$ case.

This letter is organized as follows.
In the section 2
 we introduce a new term 
to the full unitary matrix model.
We show that 
the partition function of our model
satisfy the discrete time Toda 
equation.
In the section 3
we see the role of the discrete time $l$
in the Painlev\'{e} equation.
Coupling the Toda equation and 
the string equation, 
we obtain the Painlev\'{e} V with $\delta=0$.
This becomes   the Painlev\'{e} III.
In the section 4
we consider the physical meaning of our model.
We derive the  continuum limit of the action which will
turn out to be the heat kernel 
with the theta term.
We interpret the continuum partition function 
in terms of the underlying  topology and show 
the presence of a phase transition of 
at $\theta=\pi$.
The ground state is indicated by the discrete time $l$.
The latter half of this section 
we calculate the partition function for $SU(N)$ and 
study an  effect of the theta  term.
The last section is devoted to concluding remarks.

\setzero

\section {The partition function  on the lattice}
We consider  the partition function 
of the full unitary matrix model  with  a discrete time.
We calculate explicitly  the determinant form of the partition function.

We consider the unitary matrix model 
\begin{equation}
Z_{N,l}^{U}=\int dU\exp (-\frac{N}{g^{2}}V(U)),
\end{equation}
where $V(U)$ is a potential
\begin{equation}
V(U)=t_{1}U+t_{-1}U^{\dag}-l\log U.
\label{pot}
\end{equation}
$U$ is the gauge group $U(N)$.
Here we consider the case 
\begin{equation}
\frac{t_{1}}{t_{-1}}=O(1).
\end{equation}
We parameterize $t_{1}$ and $t_{-1}$ by  $\epsilon$:
\begin{equation}
t_{1}=-e^{\epsilon},\;\;\;t_{-1}=-e^{-\epsilon}.
\end{equation}
The measure $dU$ may be written as 
\begin{equation}
dU=\prod_{m}^{N}\frac{d\alpha_{m}}{2\pi}
\Delta(\alpha)\bar{\Delta}(\alpha).
\end{equation}

 Here the eigenvalues of $U$ 
 are $\{\exp(i\alpha_{1}),\exp(i\alpha_{2}),\cdots,\exp(i\alpha_{N})\}$
and 
$\Delta\bar{\Delta}$
 is the Jacobian for the change of variables,
\begin{eqnarray}
\Delta(\alpha)&=&{\rm det}_{jk}e^{i\alpha_{j}(N-k)},
\nonumber \\ 
\bar{\Delta}(\alpha)&=&{\rm det}_{jk}e^{-i\alpha_{j}(N-k)}.
\end{eqnarray}
In terms of these variables  the potential $V(U)$ is
\begin{equation}
V(\alpha)=-\sum_{j=1}^{N}
(\cosh\epsilon\cos\alpha_{j}+i\sinh\epsilon\sin\alpha_{j}+il\alpha_{j}).
\end{equation}
Putting this all together and introducing 
\begin{equation}
f(\alpha)=
\cosh\epsilon\cos\alpha+i\sinh\epsilon\sin\alpha+il\alpha,
\end{equation}
we obtain
\begin{eqnarray}
Z_{N,l}^{U}(N/g^{2})
&=&
{\rm const.}
\prod_{m=1}^{N}\int_{-\pi}^{\pi}\frac{d\alpha_{m}}{2\pi}
e^{f(\alpha_{m})}|\Delta(\alpha)|^{2},
\nonumber \\
&=&
{\rm const.}
\prod_{m=1}^{N}\int_{-\pi}^{\pi}\frac{d\alpha_{m}}{2\pi}
e^{f(\alpha_{m})}
[\sum_{\sigma\in S_{N}}(-1)^{\sigma}\prod_{j}^{N}e^{i\alpha_{m}}(N-\sigma_{j})]
\nonumber \\
& &
\times[\sum_{\eta\in S_{N}}(-1)^{\eta}\prod_{j}^{N}e^{i\alpha_{m}}(N-\eta_{j})],
\nonumber \\
&=&
{\rm const.}
\sum_{\mu\in S_{N}}(-1)^{\mu}\sum_{\sigma\in S_{n}}
\prod_{m=1}^{N}\int_{-\pi}^{\pi}
\frac{d\alpha_{m}}{2\pi}
\nonumber \\
& & \times\exp[f(\alpha_{m})
-i\alpha_{m}(\sigma(m)-\mu\sigma(m))],
\nonumber \\
&=&
{\rm const.}
N!{\rm det}_{jk}[\int_{-\pi}^{\pi}
\frac{d\alpha}{2\pi}e^{f(\alpha)-i\alpha(k-j)}].
\label{dt}
\end{eqnarray}
where the summations are on all the permutations 
of the $N$ elements.
We can calculate the $(j,k)$ element of the determinant as 
\begin{eqnarray}
& &\int_{-\pi}^{\pi}\frac{d\alpha}{2\pi}
\exp[\frac{N}{g^{2}}(\cosh\epsilon\cos\alpha+i\sinh\epsilon\sin\alpha)
-im\alpha]
\nonumber \\
&=&
\int_{-\pi}^{\pi}\frac{d\alpha}{2\pi}
\exp[\frac{N}{g^{2}}\cos(\alpha-i\epsilon)-im\alpha]
=e^{m\epsilon}I_{m}(N/g^{2}),
\end{eqnarray}
where $m=l-j+k$.
Here $I_{m}$ is the modified Bessel function of order $m$.
Substituting this into (\ref{dt}) we  can obtain the 
partition function 
\begin{equation}
Z_{N,l}^{U}={\rm const.}
N!
{\rm det}_{jk}
e^{\epsilon(l-j+k)}I_{l-j+k}(N/g^{2}).
\label{pf21}
\end{equation}

Here we introduce symbols $D$, $D_{b}^{a}$ and $D_{b,d}^{a,c,}$ which are the 
determinants of the $(N+1)\times (N+1)$, $N\times N$ and $(N-1)\times
(N-1)$ matrices respectively with the respective definitions.
We define 
$D=Z_{N+1,l}^{U}$.
 $D_{b}^{a}$ is  same as $D$ except that the $a$-th row and 
$b$-th column are removed from it.
$D_{b,d}^{a,c}$  is same as $D$ except that the $a$-th and $c$-th
rows, and $b$-th and $d$-th columns are removed 
from it.
Then the Jacobi formula
\begin{equation}
D_{1}^{1}D_{N+1}^{N+1}-D_{1}^{N+1}\cdot D_{N+1}^{1}=
D\cdot D_{1,N+1}^{1,N+1},
\end{equation}
becomes the recurrent relations
\begin{equation}
(Z_{N,l}^{U})^{2}-Z_{N,l+1}^{U}\cdot Z_{N,l-1}^{U}
= Z_{N+1,l}^{U}\cdot Z_{N-1,l}^{U}
\label{rr}
\end{equation}
This is the discrete time Toda equation.
$l$ is the discrete time variable.
In the continuous limit of $l$, 
(\ref{rr}) becomes the Toda molecule equation.
(\ref{rr}) is the same relations 
which is  obtained from the symmetric model.\cite{gl}

\setzero
\section{Unitary Matrix Model and  Painlev\'{e} III}

It is well known that  the partition function $Z_{N}^{U}$ 
of the unitary matrix model  can be presented  as 
a product  of norms  of the biorthogonal polynomial system.
Namely, let us introduce  a scalar product of the form 
\begin{equation}
<A,B>=\oint\frac{d\mu(z)}{2\pi i z}
\exp\{-V(z)\}
A(z)B(z^{-1}),
\end{equation}
where 
\begin{equation}
V(z)=\sum_{m>0}(t_{m}z^{m}+t_{-m}z^{-m})-l\log z.
\end{equation}
Let us define  the system of the polynomials  biorthogaonal 
with respect to this scalar product 
\begin{equation}
<\Phi_{n,l},\Phi_{k,l}^{*}>=h_{k,l}\delta_{nk}.
\label{or}
\end{equation}
Then, the partition function  $Z_{N,l}^{U}$ 
is equal to the product of $h_{n}$'s:
\begin{equation}
Z_{N,l}^{U}=\prod_{k=0}^{N-1}h_{k,l},\;\;\;\tau_{0}=1.
\end{equation}
The polynomials are normalized as follows
(we should stress that superscript `*' does not mean the 
complex conjugation):
\begin{equation} 
\Phi_{n,l}=z^{n}+\cdots+S_{n-1,l},\;\;\Phi_{n,l}^{*}
=z^{n}+\cdots+S_{n-1,l}^{*},\;\;
S_{-1,l}=S_{-1,l}^{*}\equiv 1.
\label{2.4}
\end{equation}
Now it is easy to show that these polynomials satisfy the following
recurrent relations, 
\begin{eqnarray}
\Phi_{n+1,l}(z)&=&z\Phi_{n,l}(z)+S_{n,l}z^{n}\Phi_{n,l}^{*}(z^{-1}),
\nonumber \\
\Phi_{n+1,l}^{*}(z^{-1})&=&z^{-1}\Phi_{n,l}^{*}(z^{-1})
+S_{n,l}^{*}z_{-n}\Phi_{n,l}(z),
\end{eqnarray}
and
\begin{equation}
\frac{h_{n+1,l}}{h_{n,l}}=1-S_{n,l}S_{n,l}^{*}.
\end{equation}
Note that $h_{n,l}$, $S_{n,l}$, $S_{n,l}^{*}$ ,$\Phi_{n,l}(z)$
and  $\Phi_{n,l}^{*}$ depend parametrically  on
$t_{1},t_{2},\cdots,$ and $t_{-1},t_{-2},\cdots,$  but for convenience 
of notation we suppress this dependence.
Hereafter we call $t_{1},t_{2},\cdots,$ and $t_{-1},t_{-2},\cdots,$
time variables.

Using (\ref{or}) and integration by parts, we can obtain next relations:
\begin{equation}
\oint\frac{d\mu(z)}{2\pi i z} V'(z)\Phi_{n+1,l}(z)\Phi_{n,l}^{*}(z^{-1})
= (n+1)(h_{n+1,l}-h_{n,l}),
\label{se1}
\end{equation}
and 
\begin{equation}
\oint\frac{d\mu(z)}{2\pi i z} z^{2}V'(z)\Phi_{n+1,l}^{*}(z^{-1})\Phi_{n,l}(z)
= (n+1)(h_{n+1,l}-h_{n,l}).
\label{se2}
\end{equation}
(\ref{se1}) and (\ref{se2}) are string equations of 
the full unitary matrix model.

If $t_{1}$ and $t_{-1}$
are free variables  while 
$t_{2}=t_{3}=\cdots=0$ and $t_{-2}=t_{-3}=\cdots=0$,
 (\ref{se1}) and (\ref{se2})
 become
\begin{equation}
 (n+1)S_{n,l}S_{n,l}^{*}=t_{-1}(S_{n,l}S_{n+1,l}^{*}+S_{n,l}^{*}S_{n-1,l})
(1-S_{n,l}S_{n,l}^{*})-l(1-S_{n,l}S_{n,l}^{*}),
\label{edp1}
\end{equation}
\begin{equation}
 (n+1)S_{n,l}S_{n,l}^{*}=t_{1}(S_{n,l}^{*}S_{n+1,l}+S_{n,l}S_{n-1,l}^{*})
(1-S_{n,l}S_{n,l}^{*})+l(1-S_{n,l}S_{n,l}^{*}).
\label{edp2}
\end{equation}

Next we introduce a useful relation.
Using (\ref{or}) and integration by parts,  we can show
\begin{equation}
\oint\frac{d\mu(z)}{2\pi i z}zV'(z)\Phi_{n,l}(z)\Phi_{n,l}^{*}(z^{-1})
=0.
\label{1}
\end{equation}
This corresponds to  the Virasoro constraint:\cite{b}
\begin{equation}
L_{0}
=\sum_{k=-\infty}^{\infty}
kt_{k}\frac{\partial}{\partial t_{n}}.
\end{equation}
This relation constrains a symmetry like  the complex conjugate
between $t_{k}$ and $t_{-k}$.
If we set that $t_{1}$ and $t_{-1}$
are free variables  while 
$t_{2}=t_{3}=\cdots=0$ and $t_{-2}=t_{-3}=\cdots=0$,
 from (\ref{1}) we get
\begin{equation}
t_{1}S_{n,l}S_{n-1,l}^{*}=t_{-1}S_{n,l}^{*}S_{n-1,l}-l.
\label{2}
\end{equation}
Using (\ref{2}), (\ref{edp1}) and (\ref{edp2}) can be written
\begin{equation}
 (n+1)S_{n,l}=(t_{-1}S_{n+1,l}+t_{1}S_{n-1,l})
(1-S_{n,l}S_{n,l}^{*}),
\label{edp3}
\end{equation}
\begin{equation}
 (n+1)S_{n,l}^{*}=(t_{1}S_{n+1,l}^{*}+t_{-1}S_{n-1,l}^{*})
(1-S_{n,l}S_{n,l}^{*}).
\label{edp4}
\end{equation}


Using the orthogonal conditions, it is also possible to obtain the
equations which describe the time dependence of $\Phi_{n,l}(z)$ 
and $\Phi_{n,l}^{*}(z)$. 
Namely, differentiating (\ref{or}) with  respect to  times
 $t_{1}$ and $t_{-1}$ gives the following evolution equations:
\begin{equation}
\frac{\partial \Phi_{n,l}(z)}{\partial t_{1}}
=-\frac{S_{n,l}}{S_{n-1,l}}\frac{h_{n,l}}{h_{n-1,l}}
(\Phi_{n,l}(z)-z\Phi_{n-1,l}),
\end{equation}
\begin{equation}
\frac{\partial \Phi_{n,l}(z)}{\partial t_{-1}}
=\frac{h_{n,l}}{h_{n-1,l}}\Phi_{n-1,l}(z),
\end{equation}
\begin{equation}
\frac{\partial \Phi_{n,l}^{*}(z^{-1})}{\partial t_{1}}
=\frac{h_{n,l}}{h_{n-1,l}}\Phi_{n-1,l}^{*}(z^{-1}),
\end{equation}
\begin{equation}
\frac{\partial \Phi_{n,l}^{*}(z^{-1})}{\partial t_{-1}}
=-\frac{S_{n,l}^{*}}{S_{n-1,l}^{*}}\frac{h_{n,l}}{h_{n-1,l}}
(\Phi_{n,l}^{*}(z^{-1})-z^{-1}\Phi_{n-1,l}^{*}),
\end{equation}
The compatibility condition gives the following nonlinear evolution equations:
\begin{equation}
\frac{\partial S_{n,l}}{\partial t_{1}}=-S_{n+1,l}\frac{h_{n+1,l}}{h_{n,l}},
\;\;\;
\frac{\partial S_{n,l}}{\partial t_{-1}}=-S_{n-1,l}\frac{h_{n+1,l}}{h_{n,l}},
\label{11}
\end{equation}
\begin{equation}
\frac{\partial S^{*}_{n,l}}{\partial t_{1}}=S_{n+1,l}^{*}\frac{h_{n+1,l}}{h_{n,l}},
\;\;\;
\frac{\partial S^{*}_{n,l}}{\partial t_{-1}}=-S_{n-1,l}^{*}\frac{h_{n+1,l}}{h_{n,l}},
\label{22}
\end{equation}
\begin{equation}
\frac{\partial h_{n,l}}{\partial t_{1}}=S_{n,l}S_{n-1,l}^{*}h_{n,l},
\;\;\;
\frac{\partial h_{n,l}}{\partial t_{-1}}=S_{n,l}^{*}S_{n-1,l}h_{n,l},
\end{equation}
Here we define  $a_{n,l}$, $b_{n,l}$ and $b_{n,l}^{*}$:
\begin{equation}
a_{n,l}\equiv 1-S_{n,l}S_{n,l}^{*}=\frac{h_{n+1,l}}{h_{n,l}},
\end{equation}
\begin{equation}
b_{n,l}\equiv S_{n,l}S_{n-1,l}^{*},
\end{equation}
\begin{equation}
b_{n,l}^{*}\equiv S_{n,l}^{*}S_{n-1,l}.
\end{equation}
Notice that since the definitions  $a_{n,l}$, $b_{n,l}$ and $b_{n,l}^{*}$
satisfy the following identity:
\begin{equation}
b_{n,l}b_{n,l}^{*}=(1-a_{n,l})(1-a_{n-1,l}).
\label{*1}
\end{equation}
 It can be shown  using (\ref{2}) that 
\begin{equation}
t_{1}b_{n,l}=t_{-1}b_{n,l}^{*}-l.
\label{*2}
\end{equation}
In terms of  $a_{n,l}$, $b_{n,l}$ and $b_{n,l}^{*}$,
 (\ref{11}) and (\ref{22}) become the two-dimensional  Toda  equations:
\begin{equation}
\frac{\partial a_{n,l}}{\partial t_{1}}
=a_{n,l}(b_{n+1,l}-b_{n,l}),\;\;\;
\frac{\partial b_{n,l}}{\partial t_{-1}}
=a_{n,l}-a_{n-1,l},
\label{t1}
\end{equation}
and 
\begin{equation}
\frac{\partial a_{n,l}}{\partial t_{-1}}
=a_{n,l}(b_{n+1,l}^{*}-b_{n,l}^{*}),\;\;\;
\frac{\partial b_{n,l}^{*}}{\partial t_{1}}
=a_{n,l}-a_{n-1,l}.
\label{t2}
\end{equation} 

Using $a_{n,l}$, $b_{n,l}$ and $b_{n,l}^{*}$,
 we  rewrite (\ref{edp1}) and (\ref{edp2})
\begin{equation}
\frac{n+1}{t_{1}}\frac{1-a_{n,l}}{a_{n,l}}
=b_{n+1,l}+b_{n,l},
\label{s1}
\end{equation}
and 
\begin{equation}
\frac{n+1}{t_{-1}}\frac{1-a_{n,l}}{a_{n,l}}
=b_{n+1,l}^{*}+b_{n,l}^{*}.
\label{s2}
\end{equation}

From (\ref{t1}) and  (\ref{s1}) 
we eliminate $b_{n+1,l}$,
\begin{equation}
2b_{n,l}=\frac{1}{a_{n,l}}[\frac{n+1}{t_{1}}(1-a_{n,l})-
\frac{\partial a_{n,l}}{\partial t_{1}}].
\label{ww1}
\end{equation}
In the same way, from (\ref{t2}) and  (\ref{s2}) 
we eliminate $b_{n+1,l}^{*}$,
\begin{equation}
2b_{n,l}^{*}=\frac{1}{a_{n,l}}[\frac{n+1}{t_{-1}}(1-a_{n,l})-
\frac{\partial a_{n,l}}{\partial t_{-1}}].
\label{ww2}
\end{equation}
Using (\ref{*1}) and (\ref{*2}),
(\ref{t1}) and (\ref{t2}) can be written 
\begin{equation}
\frac{\partial b_{n,l}}{\partial t_{-1}}
=
(a_{n,l}-1)+\frac{1}{t_{-1}}\frac{b_{n,l}(t_{1}b_{n,l}+l)}{1-a_{n,l}},
\label{w1}
\end{equation}
\begin{equation}
\frac{\partial b_{n,l}^{*}}{\partial t_{1}}
=
(a_{n,l}-1)+\frac{1}{t_{1}}\frac{b_{n,l}^{*}(t_{-1}b_{n,l}-l)}{1-a_{n,l}}.
\label{w2}
\end{equation}
Using  (\ref{ww1}) and (\ref{w1})
to eliminate $b_{n,l}$
we obtain a second order ODE for $a_{n,l}$
\begin{eqnarray}
\frac{\partial ^{2}a_{n,l}}{\partial t_{1}\partial t_{-1}}
&=&
\frac{n+1}{t_{-1}a_{n,l}}\frac{\partial a_{n,l}}{\partial t_{1}}
-\frac{n+1}{t_{1}a_{n,l}}\frac{\partial a_{n,l}}{\partial t_{-1}}
-2a_{n,l}(a_{n,l}-1)
+\frac{(n+1)^{2}}{2t_{1}t_{-1}}
\frac{a_{n,l}-1}{a_{n,l}}
\nonumber \\
& &
+
\frac{1}{a_{n,l}}\frac{\partial a_{n,l}}{\partial t_{1}}
\frac{\partial a_{n,l}}{\partial t_{-1}}
+
\frac{1}{2}\frac{t_{1}}{t_{-1}}
\frac{1}{(a_{n,l}-1)a_{n,l}}
(\frac{\partial a_{n,l}}{\partial t_{1}})^{2}
-\frac{l^{2}}{2t_{1}t_{-1}}\frac{a_{n,l}}{a_{n,l}-1}.
\label{k1}
\end{eqnarray}
In the same way, we eliminate $b_{n,l}^{*}$
 using (\ref{ww2}) and (\ref{w1})
and obtain an ODE for $a_{n,l}$
\begin{eqnarray}
\frac{\partial ^{2}a_{n,l}}{\partial t_{1}\partial t_{-1}}
&=&
\frac{n+1}{t_{1}a_{n,l}}\frac{\partial a_{n,l}}{\partial t_{-1}}
-\frac{n+1}{t_{-1}a_{n,l}}\frac{\partial a_{n,l}}{\partial t_{1}}
-2a_{n,l}(a_{n,l}-1)
+\frac{(n+1)^{2}}{2t_{1}t_{-1}}
\frac{a_{n,l}-1}{a_{n,l}}
\nonumber \\
& &
+
\frac{1}{a_{n,l}}\frac{\partial a_{n,l}}{\partial t_{1}}
\frac{\partial a_{n,l}}{\partial t_{-1}}
+
\frac{1}{2}\frac{t_{-1}}{t_{1}}
\frac{1}{(a_{n,l}-1)a_{n,l}}
(\frac{\partial a_{n,l}}{\partial t_{-1}})^{2}
-\frac{l^{2}}{2t_{1}t_{-1}}\frac{a_{n,l}}{a_{n,l}-1}.
\label{k2}
\end{eqnarray}
Using (\ref{*2}), (\ref{t1}) and (\ref{t2}) we can obtain
\begin{equation}
t_{1}\frac{\partial a_{n,l}}{\partial t_{1}}
=
t_{-1}\frac{\partial a_{n,l}}{\partial t_{-1}}
\end{equation}
 So  $a_{n,l}$ are  functions of the radial coordinate
\begin{equation}
x=t_{1}t_{-1},
\end{equation}
only.
Then from (\ref{k1}) and (\ref{k2}) 
we can obtain
\begin{eqnarray}
\frac{\partial ^{2}a_{n,l}}{\partial x^{2}}
&=&
\frac{1}{2}(\frac{1}{a_{n,l}-1}+\frac{1}{a_{n,l}})
(\frac{\partial a_{n,l}}{\partial x})^{2}
-\frac{1}{x}\frac{\partial a_{n,l}}{\partial x}
\nonumber \\
& &
-\frac{2}{x}a_{n,l}(a_{n,l}-1)
+\frac{(n+1)^{2}}{2x^{2}}
\frac{a_{n,l}-1}{a_{n,l}}
-\frac{l^{2}}{2x^{2}}
\frac{a_{n,l}}{a_{n,l}-1}.
\label{p3}
\end{eqnarray}
$(\ref{p3})$  is an expression of the fifth Painlev\'{e}
equation (PV) with $\delta_{V}=0$
It is interesting that (\ref{p3})  has a symmetry $a_{n,l}\leftrightarrow a_{n,l}-1$
and $(n+1)^{2}\leftrightarrow -l^{2}$.
To see this, we define $c_{n,l}$:
\begin{equation}
c_{n,l}=1-a_{n,l}.
\end{equation}

Using $c_{n,l}$,
 through the transformation $x\rightarrow -x$ we rewrite $(\ref{p3})$
\begin{eqnarray}
\frac{\partial ^{2}c_{n,l}}{\partial x^{2}}
&=&
\frac{1}{2}(\frac{1}{c_{n,l}-1}+\frac{1}{c_{n,l}})
(\frac{\partial c_{n,l}}{\partial x})^{2}
-\frac{1}{x}\frac{\partial c_{n,l}}{\partial x}
\nonumber \\
& &
- \frac{2}{x}c_{n,l}(c_{n,l}-1)
+\frac{l^{2}}{2x^{2}}
\frac{c_{n,l}-1}{c_{n,l}}
-\frac{(n+1)^{2}}{2x^{2}}
\frac{c_{n,l}}{c_{n,l}-1}.
\label{p33}
\end{eqnarray}
In (\ref{p33})
the role of $l^{2}$ and  $(n+1)^{2}$ exchange.

This equation can be transformed into   the third Painlev\'{e}
 equation (P III).
(see \cite{o},\cite{h2})
(\ref{p3}) is the same equation obtained from  the O(4) nonlinear $\sigma$-model.\cite{s}.


\setzero
\section{Topology on the lattice}
We  divide   the potential  into the symmetric and the anti-symmetric
 part. 
\begin{equation}
V(U)=t_{1}^{+}s_{w}+t_{1}^{-}s_{\theta},
\end{equation}
where
\begin{equation}
t_{1}^{+}
=
\frac{t_{1}+t_{-1}}{2},\;\;\;
t_{1}^{-}
=
\frac{t_{1}-t_{-1}}{2}.
\end{equation}
$s_{w}$ is the symmetric part, the usual Wilson action
\begin{equation}
s_{w}=\frac{1}{2}({\rm tr}U+{\rm tr}U^{\dag}).
\end{equation}
Here we choose 
\begin{equation}
s_{\theta}=\frac{1}{2}({\rm tr}U-{\rm tr}U^{\dag}),
\end{equation}
for the theta term, the anti-symmetric part.
In the naive continuum  limit, 
 the sum of the theta term gives the first Chern number 
of the bundle on which the gauge field lives.  
We parametrise  the one plaquette action as 
\begin{equation}
s(U)=-\cosh \epsilon s_{w}(U)-\sinh \epsilon s_{\theta}(U) 
-l\log U,
\end{equation}
where $\epsilon$ is a real parameter that determines
 the relative weight of  the two terms, introducing in the section 2.

We  calculated the partition function in (\ref{pf21}).
We consider the continuum limit of the partition function
to leading order in $g\rightarrow 0$.
As $g\rightarrow 0$ we want the coupling of the theta term to approach 
its continuum value
$\theta/2\pi$.
Accordingly $\epsilon$ has to be scaled in this limit
as
\begin{equation}
\epsilon=\frac{\theta}{2\pi}\frac{g^{2}}{N}.
\label{cc}
\end{equation}
 The asymptotic representation of the 
modified Bessel functions are 
\begin{equation}
I_{n}(z)\longrightarrow
\frac{e^{z}}{\sqrt{2\pi z}}e^{-\frac{n^{2}}{2z}}
\;\;\;\;{\rm as} \;\;\;\; z\longrightarrow \infty.
\label{bf}
\end{equation}
Substituting (\ref{cc}) for $\epsilon$ and (\ref{bf})
into (\ref{pf21}), we obtain the small $g$
\begin{eqnarray}
Z_{l}^{U}(g,\theta)&\sim&{\rm det}_{jk}
\exp [-\frac{g^{2}}{2N}\{(l-j+k)^{2}-
\frac{\theta}{\pi}(l-j+k)\}
\nonumber \\
&=&
{\rm det}_{jk}
\exp [-\frac{g^{2}}{2N}\{(l-j+k-\frac{\theta}{2\pi})^{2}
-\frac{1}{4}(\frac{\theta}{\pi})^{2}\}].
\label{ie}
\end{eqnarray} 
Notice that $\theta$ couples to the discrete time $l$.  
For our purpose  we are interested in 
the coupling of $\theta$ and $l$.
Then in the first step 
we drop the independent  factors.
The action is given by
\begin{equation}
s(\theta,l)=-\log
 ({\rm det}_{jk}
\exp [-\frac{g^{2}}{2N}\{(l-j+k-\frac{\theta}{2\pi})^{2}
-\frac{1}{4}(\frac{\theta}{\pi})^{2}\}]).
\label{ac}
\end{equation}

If $\theta$ is varied, the ground state might change
at some points due to the level crossing 
which causes phase transition.

For $\theta=0$ using next relation
\begin{equation}
{\rm det}_{jk}\exp -\alpha(m-j+k)^{2}
=
e^{-\alpha m^{2} N}
{\rm det}_{jk}\exp -\alpha(-j+k)^{2},
\label{www}
\end{equation}
the action is
\begin{equation}
s(0,l)=-\log
 ({\rm det}_{jk}
\exp [-\frac{g^{2}}{2N}(l-j+k)])\geq
-\log
 ({\rm det}_{jk}
\exp [-\frac{g^{2}}{2N}(-j+k)])=s(0.0).
\end{equation}
It is clear that at $\theta=0$ the ground state 
is $l=0$.

For any $\theta$ using (\ref{www})
we can obtain  that
\begin{eqnarray}
s(\theta,l)&=&-\frac{g^{2}}{8}(\frac{\theta}{\pi})^{2}-\log
 {\rm det}_{jk}
\exp [-\frac{g^{2}}{2N}\{(l-j+k-\frac{\theta}{2\pi})^{2}]
\nonumber \\
&=&
-\frac{g^{2}}{8}(\frac{\theta}{\pi})^{2}
+\frac{g^{2}}{2}(l-\frac{\theta}{2\pi})^{2}
-\log
 {\rm det}_{jk}
\exp [-\frac{g^{2}}{2N}(-j+k)^{2}].
\label{fe}
\end{eqnarray}
The first term 
is a constant shift of the free energy.
The second term shows that by increasing $\theta$ staring from 
zero we do not encounter any phase transition 
 up to $\theta=\pi$.
At this point however there is a phase transition.
If $\theta$ is above $\pi$, $l=1$ becomes the ground state.
Due to the periodicity in $\theta$ the same type of phase transition 
occurs whenever $\theta$ becomes  an odd integer  times $\pi$.(Fig.1.)
{}From (\ref{fe}) the free energy density around the phase transition
is given by
\begin{eqnarray}
f(\theta)&=&0, \;\;\;\; -\pi\geq\theta\geq\pi,
\nonumber \\
f(\theta)&=&\frac{g^{2}}{2}(1-\frac{\pi}{\theta})
 \;\;\;\; -3\pi\geq\theta\geq -\pi, \,\,3\pi\geq\theta\geq \pi,
\nonumber \\
f(\theta)&=&g^{2}(2-\frac{\pi}{\theta})
 \;\;\;\; -5\pi\geq\theta\geq -3\pi,\,\, 5\pi\geq\theta\geq 3\pi,
\label{pp}
\end{eqnarray}
We have to note that (\ref{pp}) gives only the part of the energy 
without the constant  term.

Next we consider the $SU(N)$ case.
In this case there is an additional 
$\sum_{k=1}^{N}\alpha_{k}=0$ constraint on the eigenvalues
that must be enforced by a delta function in the measure.
If the delta function is written as
\begin{equation}
\delta(\sum_{k=1}^{N}\alpha_{k})
=
\sum_{m=-\infty}^{\infty}
\exp [im\sum_{k=1}^{N}\alpha_{k}],
\end{equation}
The calculation (\ref{dt}) goes through
without any modification giving the final result,
\begin{eqnarray}
Z_{N,l}^{SU}&=&{\rm const.}
N!
\sum_{m=-\infty}^{\infty}
{\rm det}_{jk}
e^{\epsilon(l+m-j+k)}I_{l+m-j+k}(N/g^{2})
\nonumber \\
&=&
{\rm const.}
N!
\sum_{m=-\infty}^{\infty}
{\rm det}_{jk}
e^{\epsilon(m-j+k)}I_{m-j+k}(N/g^{2}).
\label{pf22}
\end{eqnarray}
In the last step 
the discrete time $l$  disappears for the summing about $m$.
Using (\ref{bf}) for small $g$,
 we obtain
\begin{eqnarray}
Z_{l}^{SU}(g,\theta)&=&
\sum_{m=-\infty}^{\infty}
{\rm  const.}
N!
\sum_{m=-\infty}^{\infty}
{\rm det}_{jk}
e^{\epsilon(m-j+k)}I_{l+m-j+k}(N/g^{2})
\nonumber \\
&\sim&
\sum_{m=-\infty}^{\infty}
{\rm det}_{jk}
\exp [-\frac{g^{2}}{2N}\{(m-j+k-\frac{\theta}{2\pi})^{2}
-\frac{1}{4}(\frac{\theta}{\pi})^{2}\}]
\nonumber \\
&=&
\sum_{m=-\infty}^{\infty}
\exp[-\frac{g^{2}}{2}(m-\frac{\theta}{2\pi})^{2}]
{\rm det}_{jk}
\exp[-\frac{g^{2}}{2N}\{(-j+k)^{2}
-\frac{1}{4}(\frac{\theta}{\pi})^{2}\}].
\label{ie1}
\end{eqnarray} 
In the $g\rightarrow 0$ limit
this sum can be explicitly evaluated  by the 
approximating it with a Gaussian integral.
Then we can obtain
\begin{equation}
Z_{l}^{SU}(g,\theta)\sim
\frac{\sqrt{2\pi}}{g}
\exp
\frac{g^{2}}{8}(\frac{\theta}{\pi})^{2}
{\rm det}_{jk}
\exp[-\frac{g^{2}}{2N}\{(j-k)^{2}
-\frac{1}{4}(\frac{\theta}{\pi})^{2}\}].
\end{equation}
Then in the $SU(N)$ case $\theta$ 
 has influence on  a constant shift of the free energy only.
We can  rewritten  (\ref{pf22}) as
\begin{eqnarray}
Z_{N,l}^{SU}&=&{\rm const.}
N!
\sum_{l=-\infty}^{\infty}
{\rm det}_{jk}
e^{\epsilon(l-j+k)}I_{l-j+k}(N/g^{2})
\nonumber \\
&=&
\sum_{l=-\infty}^{\infty}Z_{N,l}^{U}.
\end{eqnarray}
The partition function for  the $SU(N)$ case  is the 
sum of  the partition function for  the $U(N)$ case 
about all the discrete time $l$.
Thus the theta term would have been zero for $SU(N)$ case.

\setzero
\section{Concluding remarks}

In this letter 
we introduce a new term to the full unitary matrix model.
This term is $l\log U$.
We show that the partition function of this model 
satisfy the discrete time Toda equation
 and $l$ is the discrete time.
Furthermore we consider the relation the partition function of this model  
and Painlev\'{e} III.
We can see the symmetry between $n$ and $l$.

We derive the continuum limit of this model in the presence of a 
topological term.
The theta term 
remained  to be coupled to the dynamics  
in the continuum,
 giving rise to phase transition 
at $\theta_{c}=\pm\pi,\pm3\pi,\pm5\pi,\cdots$.
The ground state 
is indicated by the discrete time $l$.
The discrete time plays like  the instanton number.
The partition function for  the $SU(N)$ case  is the 
sum of  the partition function for  the $U(N)$ case 
about all the discrete time $l$.
Then  for the $SU(N)$ case the theta term 
has influence on a constant shift 
of the free energy only.

\newpage

\begin{figure}[hbtp]
\epsfile{file=fe.eps,scale=0.8}
\caption{ The free energy density around the phase transition.}
\end{figure}
\end{document}